\documentclass[journal=jctcce,manuscript=article]{achemso}
\usepackage{amsmath}
\usepackage{nameref}
\usepackage{blindtext}
\usepackage[dvipsnames]{xcolor}
\usepackage[utf8]{inputenc}
\usepackage{graphicx} 
\usepackage{bm}	
\usepackage{color}                     
\usepackage{xcolor}
\usepackage{epsfig}
\usepackage{amsmath} 
\usepackage{amssymb} 
\usepackage{longtable} 
\usepackage{float}
\usepackage{mathtools}
\usepackage{xparse}
\usepackage{hyperref}
\usepackage{times}
\usepackage[normalem]{ulem}
\usepackage{sidecap}
\usepackage{subcaption}
\usepackage{appendix} 
\usepackage{multirow}
\sidecaptionvpos{figure}{t}
\usepackage{comment}
\usepackage{tikz,forest}
\usetikzlibrary{arrows.meta}
\usepackage[ruled,vlined]{algorithm2e}
\usepackage[version=4]{mhchem}

\setkeys{acs}{etalmode=truncate,maxauthors=10}
\setkeys{acs}{articletitle=true}

\title{On the basis set selection for molecular core-level $GW$ calculations}

\author{Daniel Mejia-Rodriguez}
\email{daniel.mejia@pnnl.gov}
\affiliation{Environmental Molecular Sciences Laboratory, Pacific Northwest National Laboratory, Richland, WA 99352, USA}

\author{Alexander Kunitsa}
\email{aakunitsa@gmail.com}
\affiliation{Zapata Computing, Inc., 100 Federal Street,
Boston, MA 02110, USA}

\author{Edoardo Aprà}
\email{edoardo.apra@pnnl.gov}
\affiliation{Environmental Molecular Sciences Laboratory, Pacific Northwest National Laboratory, Richland, WA 99352, USA}

\author{Niranjan Govind}
\email{niri.govind@pnnl.gov}
\affiliation{Physical and Computational Sciences Directorate, Pacific Northwest National Laboratory, Richland, WA 99352, USA}

\begin{document}
\maketitle
\newpage
\begin{abstract}
The $GW$ approximation has been recently gaining popularity among
the method for simulating 
molecular core-level X-ray photoemission spectra.
Traditionally, $GW$ core-level binding energies have been
computed using either the cc-pV$n$Z or def2-$n$ZVP basis set
families, extrapolating the obtained results to the complete basis 
set limit, followed by a an element-specific relativistic 
correction. Despite of achieving good accuracy, these
binding energies are chronically underestimated. By using
first-row elements and standard techniques known to offer
good cost-accuracy ratio in other theories, we show that
the cc-pV$n$Z and def2-$n$ZVP families show large contraction
errors and lead to unreliable complete basis set extrapolations.
On the other hand, we demonstrate that uncontracted versions of these
basis sets offer vastly improved convergence. Even faster
convergence can be obtained using core-rich, property-optimized, 
basis sets families like pcSseg-$n$, pcJ-$n$ and ccX-$n$Z.
Finally, we also show that the improvement over the core properties does not degrade
the calculation of the valence excitations, and thus offer a 
balanced description of both core and valence regions.

\end{abstract}

\section{Introduction}
\label{intr}
X-ray-based spectroscopies are well suited to study
the local environment of atoms in a molecule or material.
Typically, a core electron is probed through electronic excitation or
ionization, leading to a very sensitive and selective 
spectrum \cite{Norman:2018:7208}. The use of state-of-the-art 
X-ray light sources has expanded the applicability and
power of such techniques. 
Moreover, it has been noted that the interpretation of the X-ray spectra requires high-level theoretical approaches\cite{Milne:2014:44,Halbert:2021:3583}. 
Theoretical simulations of X-ray-based spectroscopies are thus
increasingly relevant.

X-ray photoelectron spectroscopy (XPS)\cite{Siegbahn:1967} is one of 
the most widely used X-ray-based approaches. The key quantities in 
simulations of XPS spectra are core-level binding energies (CLBEs), which
correspond to the ionization energies in core orbitals. The most popular
methods for computing CLBEs can be classified into two main categories:
$\Delta$self-consistent-field-like ($\Delta$SCF) methods, and
response methods (including linear response, equation-of-motion, and
Green's function approaches).  

The $GW$ approximation (GWA)\cite{Hedin:1965:A796} to the
self-energy is a Green's-function-based method that can be
used to obtain accurate molecular CLBEs at a reasonable cost \cite{vanSetten:2018:877,Golze:2018:4856,Golze:2020:1840,MejiaRodriguez:2021:7504}.
The CLBEs obtained with this approach include some level
of orbital relaxation effects, however, the quality of the
results might depend strongly on the starting point. For example,
it has been known that a large fraction of exact exchange is
needed at the one-shot $G_0W_0$ level, mainly due to a large
self-interaction error (SIE) present when standard ``pure'' 
exchange-correlation density functional approximations (DFAs) are adopted.
\cite{vanSetten:2018:877,Golze:2018:4856,MejiaRodriguez:2021:7504}. 
A possible workaround when starting from a pure Kohn-Sham result
is to use the more demanding ev$GW$ partial self-consistent approach  \cite{Golze:2018:4856,Golze:2020:1840,Keller:2020:114110,MejiaRodriguez:2021:7504}, although some starting-point dependency will remain.

Besides these unique requirements, CLBEs obtained with the
GWA will also depend on the quality of the basis set.
The accuracy of GWA-based CLBEs has been studied recently by means of
two benchmark datasets \cite{vanSetten:2018:877,Golze:2020:1840,Keller:2020:114110}. Both studies extrapolated the CLBEs to the complete basis-set (CBS) limit
in order to eliminate the dependency on the basis set.
In Ref. \citenum{vanSetten:2018:877},
the authors used the def2-TZVP and def2-QZVP \cite{Weigend:2005:3297}
basis sets and omitted relativistic corrections; while in
Refs. \citenum{Golze:2018:4856} and \citenum{Golze:2020:1840},
the authors used Dunning's correlation-consistent basis sets
cc-pV$n$Z\cite{Dunning:1989:1007}, from triple- to sextuple-zeta 
quality, and included relativistic corrections. Both studies 
showed that $GW$ can be used to predict CLBEs with a mean absolute
error of about 0.3 eV for first row elements, but that this error
can only be ``inconsistently'' reduced by adding
relativistic corrections\cite{Keller:2020:114110}.

Both benchmark studies also relied on the extrapolation of the CLBEs, obtained
with standard basis sets, to the CBS limit. Standard CBS extrapolation
techniques were developed to extrapolate the correlation energy 
of small molecules \cite{Feller:1992:6104,Feller:1993:7059,Helgaker:1997:9639}.  This is usually a small-energy regime with well-known basis set convergence behavior\cite{Schwartz:1962:1015,Kutzelnigg:1992:4484}, and asymptotic behavior--$E_c/Z \sim -0.02073 \ln{Z} + 0.0372$\cite{Burke2016,Cancio2018}--for atoms. In contrast, the energy involved for ionizing the core is rather large--$\varepsilon_{1s} \sim Z^2/2$--and does not follow the same convergence trend. This may explain why Golze and co-workers used a
a linear regression fit with respect to the inverse number of basis functions \cite{Golze:2020:1840} compared with Bruneval and co-workers, who used the more common exponential convergence formula for $GW$ valence binding energies
\cite{Hung:2017:2135,Bruneval:2020:4399}.
%

Even when the correlation energy
of molecules and the core ionization energy have similar asymptotic
behaviors, the basis set requirements for a response method,
like the GWA, might be very different compared with a single-point
energy calculation. It has been shown that basis sets optimized for energies
often fail to describe other molecular properties due to a poor
description of the orbitals in regions important for the given
property but otherwise energetically unimportant \cite{Jensen:2015:132,Ambroise:2021}.
This numerical issue prevents the application of 
systematic improvements and of reliable extrapolation approaches.
A natural consequence is
that extrapolation using non-optimal basis 
sets might result in values far away from the real CBS limit.

The importance on the selection of the basis set have already
been addressed for other response methods, like the equation-of-motion
coupled-cluster \cite{Sarangi:2020:e1769872,Ambroise:2021}
 or the linear-response time-dependent density functional theory
 formalism \cite{Fouda:2018:6}. Basis set convergence has also been 
 studied within the GWA applied to molecules, mostly focused 
 to valence ionization potentials \cite{Hung:2017:2135,Bruneval:2020:4399,Loos:2020:1018},
 usually showing smooth, albeit slow, convergence. However, a systematic study comparing
 CLBEs obtained with basis sets other than the cc-pV$n$Z and def2-$n$ZVP families
is not available in the literature. 

In this study, we use the CORE65 dataset, comprised of
65 CLBEs of first-row elements, in order to test the
effect of the basis set core-level $GW$ calculations.
We asses two types of basis sets that are known to offer very good
results in other theories: uncontracted versions
of traditional energy-optimized basis sets, and core-rich 
property-optimized basis sets. 

In turn, we show that both cc-pV$n$Z and def2 basis set families 
are not well suited to describe first-row 1s CLBEs within the GWA,
often resulting in very slow convergence and unreliable CBS
extrapolations.
Findings of this work are consistent with earlier studies dealing with
other linear-response methods, by showing that while uncontracted versions
of these basis sets offer vastly improved results,
the CLBEs seem to converge to a different value as compared to
the CLBEs obtained with other core-rich
basis sets like pcJ-$n$ \cite{Jensen:2006:1360}, 
pcSseg-$n$ \cite{Jensen:2015:132}, 
and the very recent ccX-$n$Z\cite{Ambroise:2021} families. 
The difference is small (around 100 meV), however, 
this value is close to the intrinsic accuracy
of the GWA for CLBEs and, as such, it is worth
noticing.

\section{Theory}
\label{theor}

\subsection{Overview of the \texorpdfstring{$GW$}{GW} Approximation}
The central object of the GWA is the one-particle Green's function $G$
describing particle and hole scattering in the interacting many-body system.
In order to obtain the electron addition and removal energies from such a
Green's function, a non-local and dynamic effective potential, the
self-energy $\Sigma$, is often introduced. The self-energy $\Sigma$
substitutes the mean-field exchange-correlation operator, and the $GW$
quasiparticle (QP) energies $\varepsilon_n^{GW}$ can be obtained as
corrections to the mean-field energies $\varepsilon_n$:
\begin{equation}
    \varepsilon_{n\sigma}^{GW} = \varepsilon_{n\sigma} + \Re\left( \Sigma_{n\sigma}(\varepsilon_{n\sigma}^{GW}) \right) - V_{n\sigma}^{xc}
    \label{eq:qpeq}
\end{equation}

Here, $\sigma$ denotes the spin index, and $V_{n\sigma}^{xc}$ and $\Sigma_{n\sigma}$ denote the $n$th diagonal element of the corresponding matrix representation in the molecular orbital basis. Equation \ref{eq:qpeq} is non-linear and must be solved iteratively.

The self-energy operator $\Sigma$ is given in terms of the Green's function $G^{\sigma}$ and the screened Coulomb interaction $W$:
\begin{equation}
    \Sigma_\sigma(\mathbf{r},\mathbf{r}',\omega) = \frac{i}{2\pi} \int \mathrm{d}\xi G^{\sigma}(\mathbf{r},\mathbf{r}',\omega+\xi) W(\mathbf{r},\mathbf{r}',\xi) e^{i\xi\eta}
\end{equation}
with $\eta$ being a positive infinitesimal. In practice, the GWA is often performed as a one-shot perturbative approach known as $G_0W_0$. In this case, $G_0^{\sigma}$ is the non-interacting mean-field Green's function,
\begin{equation}
    G_0^\sigma(\mathbf{r},\mathbf{r}',\omega) = \sum\limits_m \frac{\phi_{m\sigma}(\mathbf{r})\phi_{m\sigma}(\mathbf{r}')}{\omega - \varepsilon_{m\sigma} - i\eta \; \mathrm{sign}(\mu - \varepsilon_{m\sigma})} \;\; ,
\end{equation}
and $W_0$ is obtained using the random phase approximation (RPA) as
\begin{equation}
    W_0(\mathbf{r},\mathbf{r}',\omega) = \int \mathrm{d}\mathbf{r}'' \epsilon^{-1} (\mathbf{r},\mathbf{r}'',\omega) v(\mathbf{r}'',\mathbf{r}')
\end{equation}
In the preceding equations, $\mu$ is the Fermi-level of the system, $v(\mathbf{r},\mathbf{r}')$ is the bare Coulomb interaction, and $\epsilon(\mathbf{r},\mathbf{r}'',\omega)$ is the RPA dynamical dielectric function:
\begin{equation}
    \epsilon(\mathbf{r},\mathbf{r}'',\omega) = \delta(\mathbf{r},\mathbf{r}') - \int \mathrm{d}\mathbf{r}'' v(\mathbf{r},\mathbf{r}'')\chi_0(\mathbf{r}'',\mathbf{r}',\omega)
\end{equation}
In the RPA, the irreducible polarizability $\chi_0$ has a simple sum-over-states representation\cite{Adler:1962,Wiser:1963}
\begin{equation}
    \chi_0(\mathbf{r},\mathbf{r}',\omega) = \sum\limits_\sigma \sum\limits_{i,a}  \left[ \frac{\phi_{i\sigma}(\mathbf{r})\phi_{a\sigma}(\mathbf{r})\phi_{i\sigma}(\mathbf{r}')\phi_{a\sigma}(\mathbf{r}')}{\omega - \varepsilon_{a\sigma} + \varepsilon_{i\sigma} +i\eta} + \frac{\phi_{i\sigma}(\mathbf{r})\phi_{a\sigma}(\mathbf{r})\phi_{i\sigma}(\mathbf{r}')\phi_{a\sigma}(\mathbf{r}')}{-\omega - \varepsilon_{a\sigma} + \varepsilon_{i\sigma} + i\eta} \right]
\end{equation}
where the index $i$ runs over the occupied orbitals while the index $a$ runs over the virtual ones. 

The contour deformation (CD) technique is used in this work for the accurate integration of $\Sigma$, as it has been shown to have a good cost-effective profile for the evaluation of CLBEs. Further details about CD-$GW$ and its implementation using local orbitals can be obtained from References \nocite{Golze:2018:4856,Holzer:2019:204116,MejiaRodriguez:2021:7504} \citenum{Golze:2018:4856}, \citenum{Holzer:2019:204116}, and \citenum{MejiaRodriguez:2021:7504} for example.

\section{Computational Details}
\label{comp}
The adequacy of the cc-pV$n$Z and def2-$n$ZVP basis set families for
core-level $GW$ calculations was evaluated by computing the 65 CLBEs 
from the CORE65 benchmark dataset\cite{Golze:2020:1840}. 
These energies were then compared with those obtained with the
respective completely uncontracted versions, hereafter denoted as
un-cc-pV$n$Z and un-def2-$n$ZVP.
The CD-$GW$ approach, recently implemented in the open-source 
computational chemistry package \textsc{NWChem}
\cite{NWChem,MejiaRodriguez:2021:7504}, was used for this task. The frequency integral over the
imaginary axis was evaluated with a modified Gauss-Legendre grid with 200
points. The QP equation were always solved iteratively (i.e. no linearized
approximation was used). All calculations used the \textsc{Simint} library
\cite{simint:2016,simint:2018} and a $10^{-14}$ Schwarz screening threshold for the
evaluation of the electron repulsion integrals. 
All CLBEs were obtained  using the PBEh \cite{Atalla:2013:165122} 
functional with 45 \% of exact exchange as starting point for the  $G_0W_0$ calculations \cite{Golze:2018:4856}.

In order to find other appropriate basis sets families, the
65 CLBEs were also obtained with the pcJ-$n$,  
pcSseg-$n$, and ccX-$n$Z families, as well as their
uncontracted versions.

Since our CD-$GW$ implementation fits the four-center ERIs, 
an adequate fitting basis is needed. Here, we do not directly
assess the impact of various fitting basis sets on the quality of
CLBEs but we do use automatically generated fitting bases
using the ``AutoAux'' procedure \cite{Stoychev:2017:554} for
each orbital basis set.

\section{Results and Discussion}
\label{resul}
\subsection{The cc-pV\emph{n}Z and def2-\emph{n}ZVP families}
\begin{figure}[ht]
    \centering
    \includegraphics[width=0.9\textwidth]{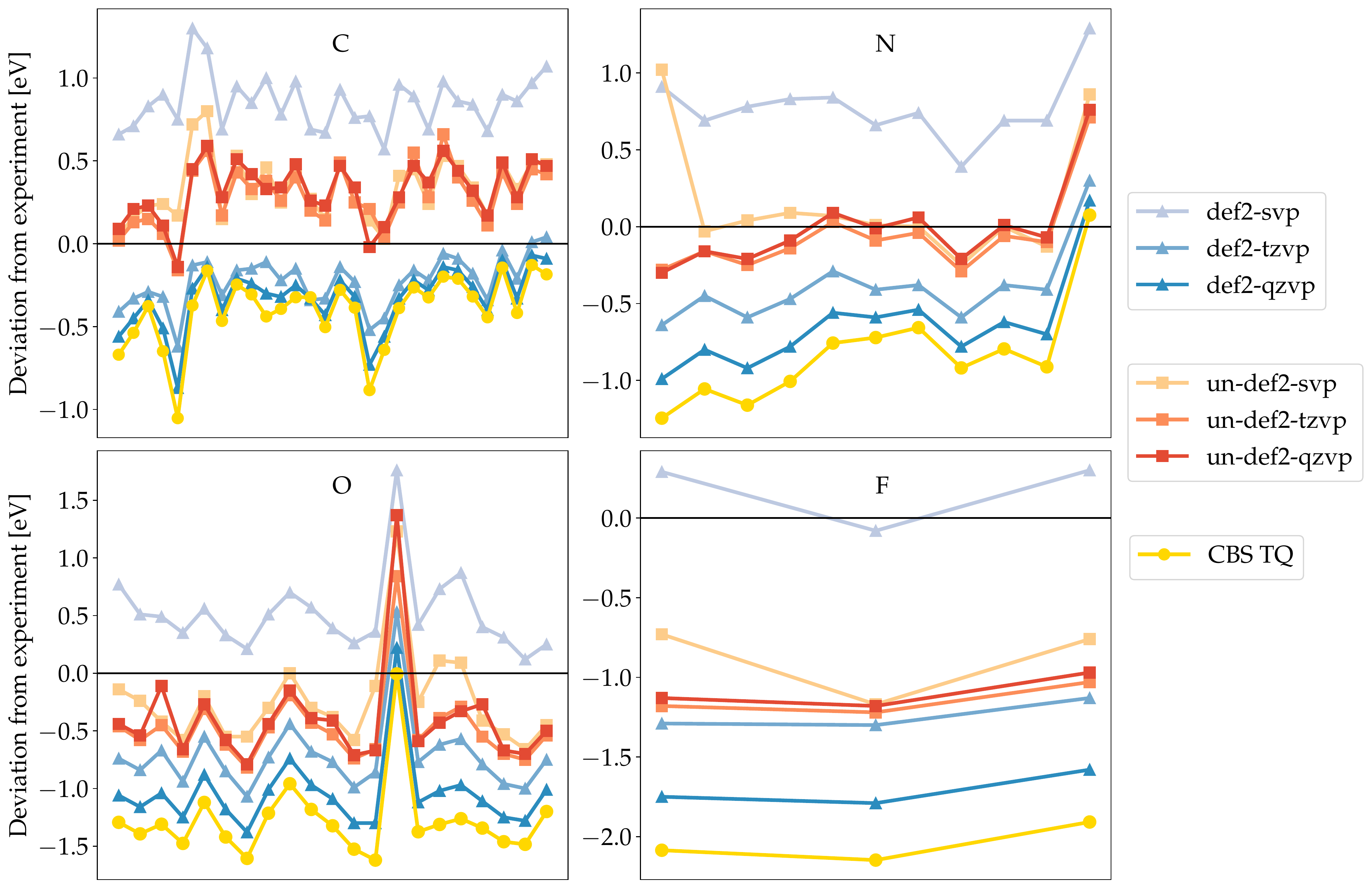}
    \caption{$G_0W_0$ 1s binding energy error due to the contraction of the def2 basis set family. The plots show shifts with respect to the experimental results listed in Reference \citenum{Golze:2020:1840}. No relativistic corrections were included. }
    \label{fig:def2}
\end{figure}

\begin{figure}[ht]
    \centering
    \includegraphics[width=0.9\textwidth]{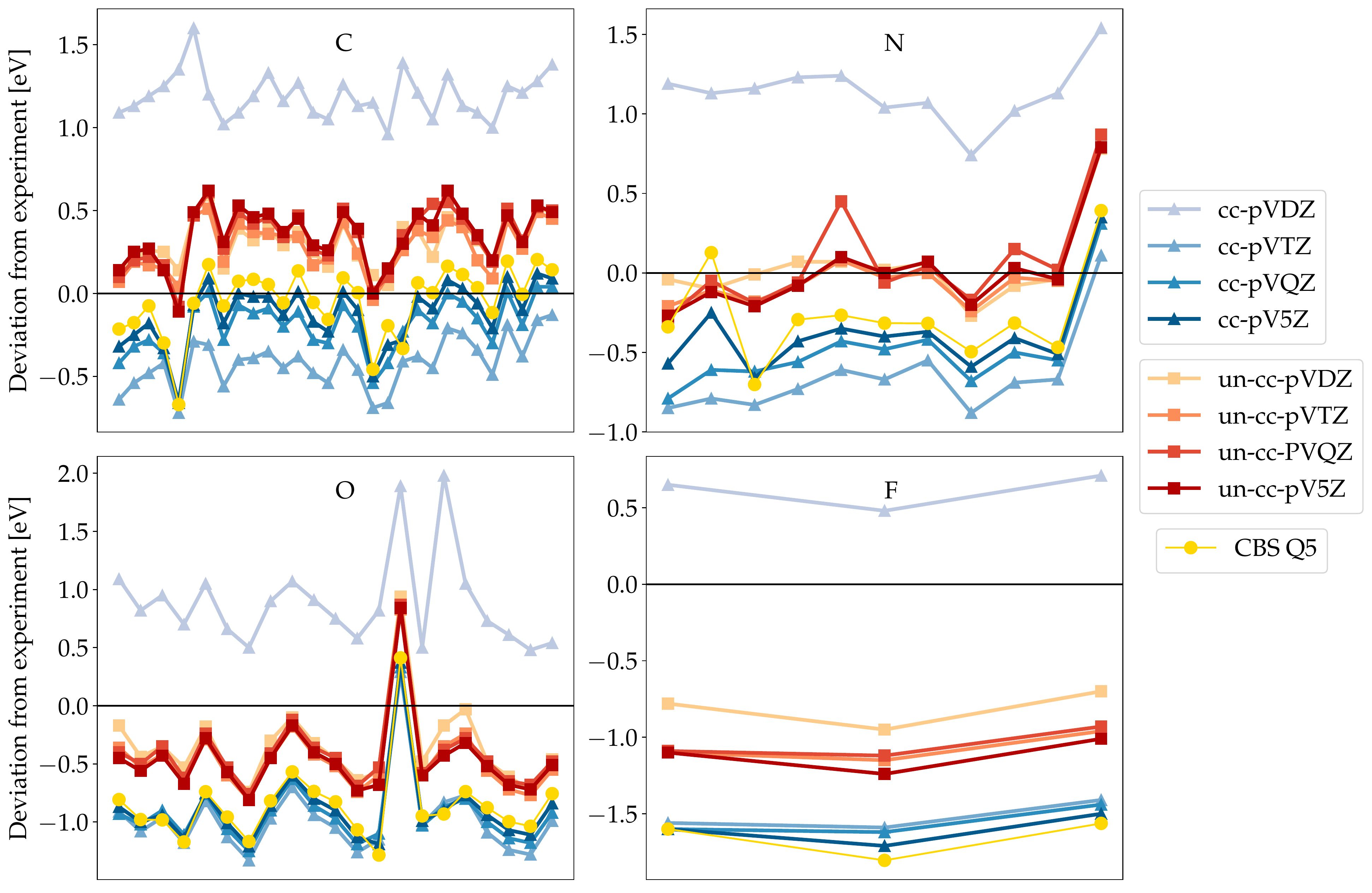}
    \caption{$G_0W_0$ 1s binding energy error due to the contraction of the cc-pV$n$Z basis set family. The plots show shifts with respect to the experimental results listed in Reference \citenum{Golze:2020:1840}. No relativistic corrections were included. }
    \label{fig:ccpvnz}
\end{figure}

We first begin studying the two families used in the CLBEs benchmark
studies in the literature, the cc-pV$n$Z and def2-$n$ZVP families, respectively.
Both basis set families were developed with total energies in mind. 
In energy-optimized basis sets, contraction of the core basis functions 
is usually performed in order to achieve high efficiency at the price 
of a small loss in accuracy \cite{Jensen:2008:719}. The removal of a core
electron, however, induces a significant relaxation of the core orbital
incompatible with a fixed shape due to the contraction scheme. 
The contraction effect is clearly visible when comparing the results of a
given basis set and its corresponding uncontracted version. 
Such an effect can be seen in Figures \ref{fig:def2} and \ref{fig:ccpvnz}
for the def2-$n$ZVP and cc-pV$n$Z families, respectively, where 
CLBEs deviations with respect to the experiment (see 
Supplemental Material of reference \citenum{Golze:2020:1840}) are shown. 
The contracted series are presented in a
blue multi-hue color scheme, while the uncontracted series in a
red multi-hue color scheme. A two-point CBS extrapolation is also
shown in gold with dots.

There are several important remarks about the plots shown
in Figures \ref{fig:def2} and \ref{fig:ccpvnz}: first, 
contracted double-$\zeta$ basis sets lead to very large 
CLBEs regardless of the element; second, CLBEs obtained with
uncontracted basis sets are almost always larger than those obtained
with contracted basis sets, the only exception being
the unusually large CLBEs obtained with def2-SVP and cc-pVDZ; 
third, since def2-QZVP CLBEs are smaller than
def2-TZVP CLBEs, a CBS extrapolation shifts the CLBEs 
prediction further away from experimental values; fourth,
CLBEs obtained with uncontracted basis sets show a rather fast
convergence, and a double-$\zeta$ basis set suffices for C, N, and O
in most cases.

These results provide evidence that the contraction schemes
used to generate the def2-$n$ZVP and cc-pV$n$Z families 
are not well suited to describe CLBEs within the GWA.
As a consequence, the CBS extrapolations used in previous
studies do not convey absolute CLBEs converged with respect to
the basis set size, often being 0.5 eV too small
(when comparing the results from uncontracted basis with the ones from CBS extrapolation).

\begin{figure}[ht]
    \centering
    \includegraphics[width=0.9\textwidth]{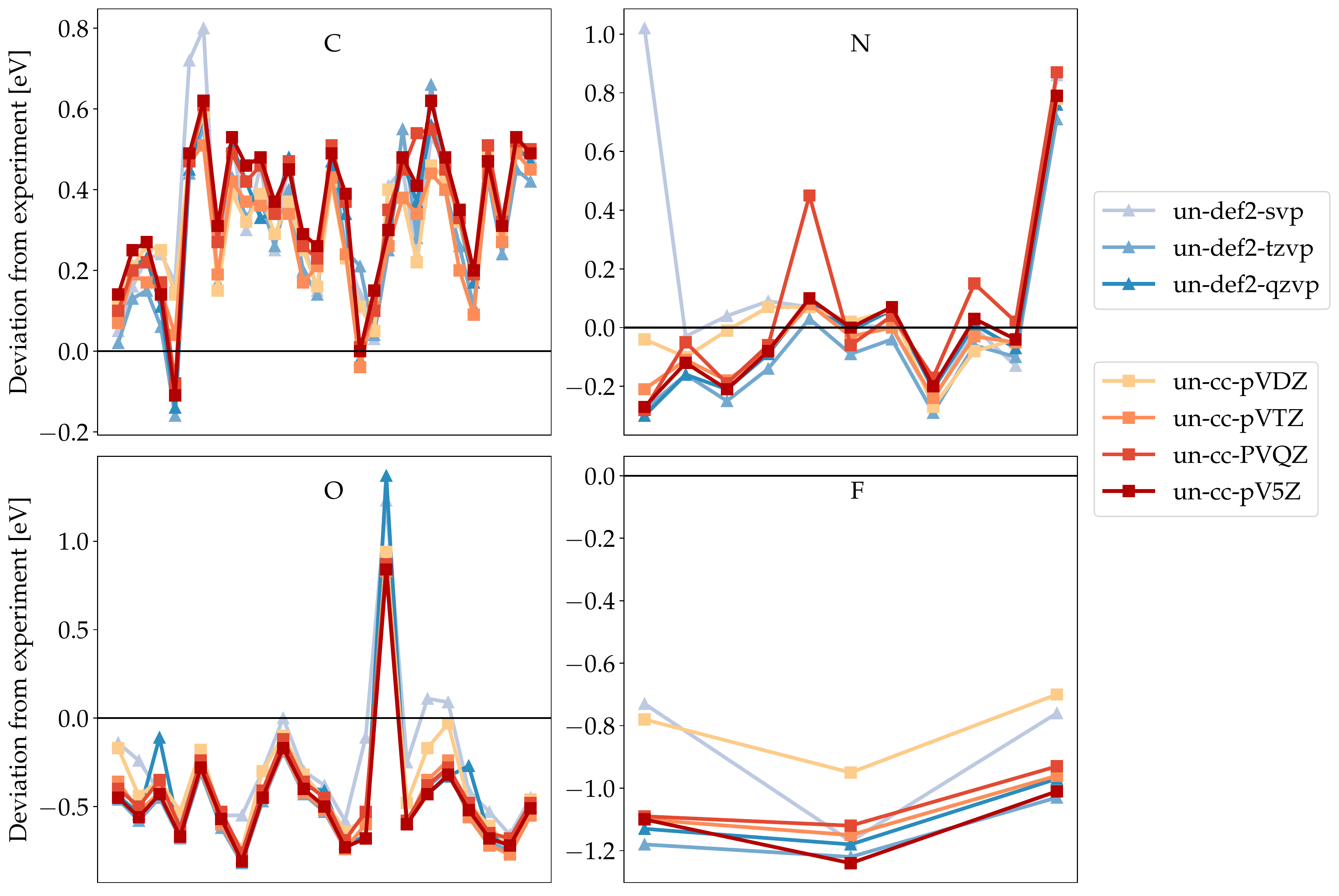}
    \caption{Comparison of $G_0W_0$ 1s binding energies obtained with
    two families of uncontracted basis sets. No relativistic corrections were included.}
    \label{fig:uncon}
\end{figure}

An arguably better description can be obtained by
using the uncontracted versions of both basis set
families. Figure \ref{fig:uncon} shows that the
CLBEs predicted by both uncontracted basis set
families are rather similar. This supports the 
notion that this description might be indeed close to the
basis set limit. It is important to mention that the CLBEs obtained with 
the uncontracted bases are systematically larger than 
the CBS extrapolation used in Reference \citenum{Golze:2020:1840} 
by about 0.25 eV for carbon and nitrogen, 
0.40 eV for oxygen, and 0.50 eV for fluorine. 

\subsection{Comparison with core-rich basis sets}
\subsubsection{Absolute core-level binding energies}
Given that the flexibility around the core seems to play a very
important role in determining the quality of the $GW$ CLBEs,
the use of core-rich basis sets might be a better choice  
than the use of either un-cc-pV$n$Z or un-def2-$n$ZVP \cite{Ambroise:2021}. 
As mentioned earlier, we could not find a systematic study of $GW$ CLBEs using these basis sets. The only data available was found in the
Supplemental Material of reference \cite{Keller:2020:114110}, where the
CLBEs obtained with the core-augmented cc-pCV$n$Z basis set was 
compared with CLBEs obtained with the standard cc-pV$n$Z basis sets. 
In that study, the authors noted that the errors of the cc-pCV$n$Z 
were not systematic. A similar conclusion was reached in 
Reference \citenum{Ambroise:2021}, where the authors noted that 
uncontracting the cc-pCV$n$Z family leads to larger deviations 
with respect to the CBS limit, a result which suggests an 
uncontrolled error cancellation in the contraction scheme.

\begin{figure}[ht]
    \centering
    \includegraphics[width=0.9\textwidth]{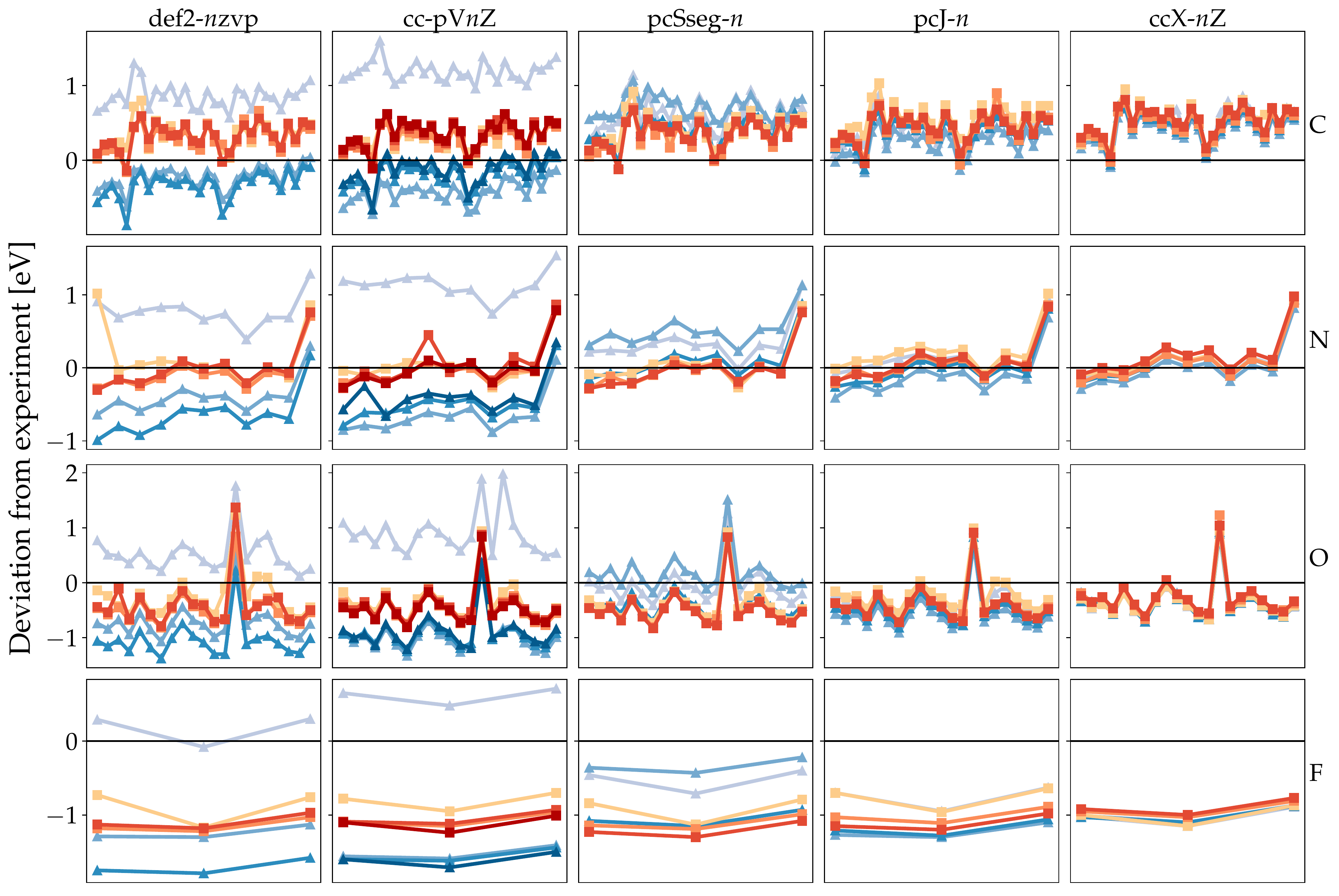}
    \caption{$G_0W_0$ 1s binding energy deviations with respect to experiment. No relativistic corrections were included.
    All families include double-, triple-, and quadruple-$\zeta$
    bases, and the cc-p$n$VZ family includes cc-pV5Z.
    The color code is analog to Figures \ref{fig:def2}--\ref{fig:uncon},
    blue hues for contracted bases and red hues for uncontracted ones.
    The larger the $n$, the darker the color.}
    \label{fig:coreaug}
\end{figure}

Here, we have extended the results presented in Reference
\citenum{Keller:2020:114110}
and have included the pcJ-$n$, pcSseg-$n$, and ccX-$n$Z core-rich basis set
families. Figure \ref{fig:coreaug} shows CLBEs deviations with respect to
experiment for all basis set families tested. The same color code as in
the previous sections is being used: blue hues for contracted sets, and 
red hues for uncontracted sets.
Note that relativistic corrections have not been included. 

The three core-rich basis set families predict converged CLBEs
in very good agreement to the ones obtained using the
uncontracted def2-$n$ZVP or cc-pV$n$Z basis sets. Moreover, 
there is no longer a mismatch between the CLBEs obtained with
contracted and uncontracted versions of these basis sets.
The fact that the same limit is achieved by all uncontracted
basis sets, as well as with the contracted pcJ-$n$, pcSseg-$n$, 
and ccX-$n$Z, strongly supports the idea that this limit is
in fact closer to the CBS than the extrapolations given in
Reference \citenum{Golze:2020:1840}.

Another aspect worth noting is the manner in which the
CLBEs converge for each basis set. As noted before, the
def2-$n$ZVP basis sets show the wrong shift going from
triple- to quadruple-$\zeta$, the cc-pV$n$Z family converges
very slowly from below, the pcSseg-$n$ family converges from
above, the pcJ-$n$ family converges from below, and the
ccX-$n$Z family is very well converged even with
double-$\zeta$ quality. 

As a consequence of the overall larger
CLBEs predicted with core-rich basis sets, the statistics presented
by Golze et al.\cite{Golze:2020:1840} also shift.
All individual CLBEs, as well as the mean errors and
mean absolute errors with respect to experiment, are
tabulated in the supplemental material. Taking 
the results obtained with the uncontracted ccX-QZ basis set
as our approximation to the CBS limit, we find that
$G_0W_0$@PBEh($\alpha=0.45$) yields mean absolute
errors of 0.65, 0.41. 0.21, and 0.19 eV for C, N, O, and
F, respectively. In comparison, the values reported
by Golze et al.\cite{Golze:2020:1840} are 0.24, 0.16,
0.48, and 0.83 eV, respectively (see Tables S11 and S12
of the Supplementary material for more details).

\subsubsection{Relative core-level binding energies}
\begin{figure}[ht]
    \centering
    \includegraphics[width=0.9\textwidth]{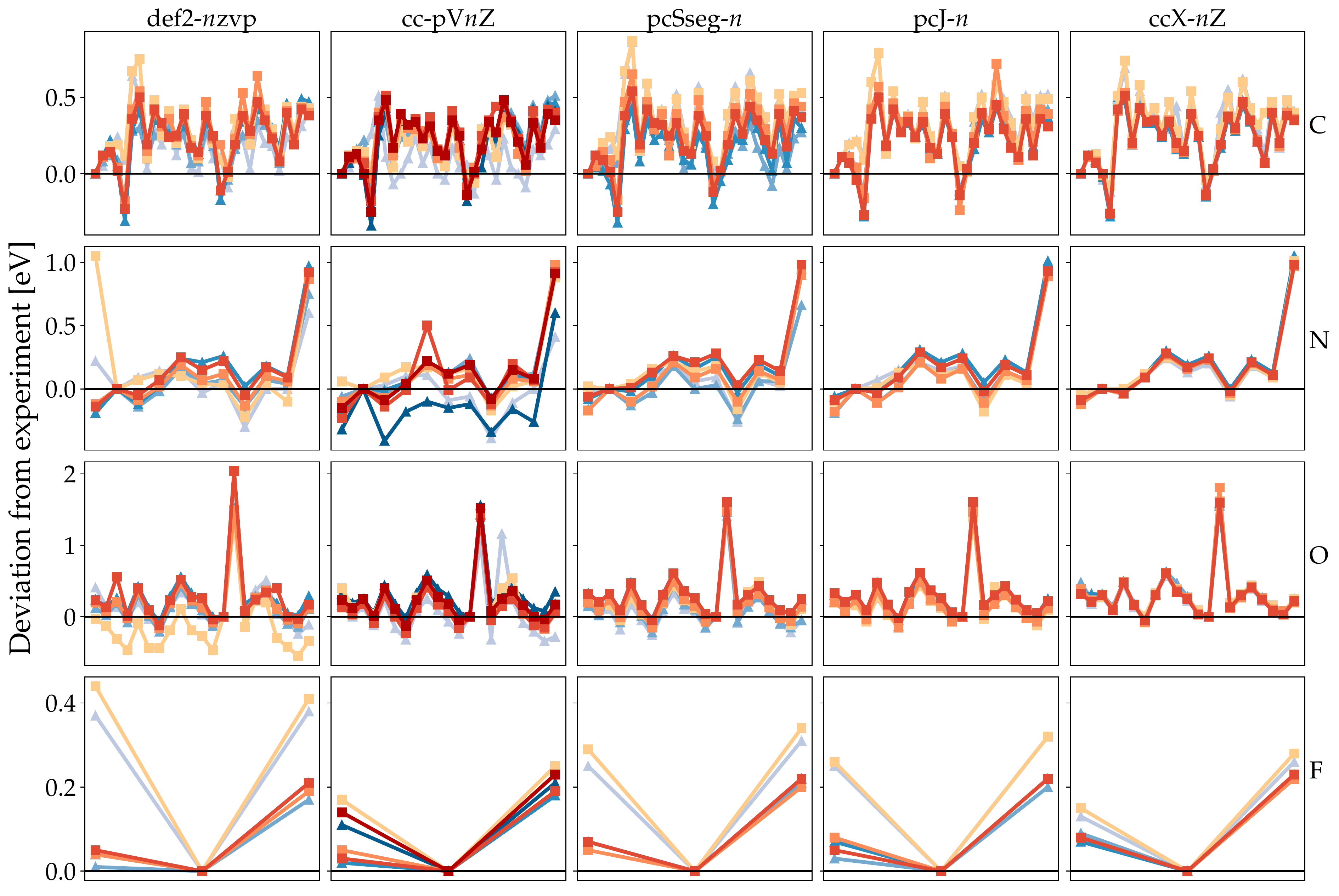}
    \caption{$G_0W_0$ 1s chemical shift deviations with respect to experiment. No relativistic corrections were included. Same coloring
    scheme as in Figure \ref{fig:coreaug}.}
    \label{fig:coreaug_shift}
\end{figure}

Chemical shifts show faster convergence with any of the basis set
used in this study, as shown in Figure \ref{fig:coreaug_shift}.
The contracted pcJ-$n$ and ccX-$n$Z tend to show 
smaller dispersions overall, however, this effect is 
rather small and barely noticeable. 
A special case is obtained with cc-pV5Z for Nitrogen 1s 
chemical shifts, as Figure \ref{fig:coreaug_shift} shows that
these chemical shifts are too small as compared to the consensus
of the remaining bases. We have not been able to identify the
cause for this discrepancy, but note that the chemical shifts
obtained with the uncontracted version are in very good agreement
with the rest of the results.

Since the relative binding energies are less sensitive to the
basis set chosen, the statistics presented by 
Golze et al.\cite{Golze:2020:1840} does not change by more than
0.05 eV when compared to our ccX-QZ results.

\subsubsection{Valence binding energies}
Improvement in convergence of core-level binding
energies should not, in principle, come at the expense
of a poor description of the valence region. We evaluated
the accuracy achieved with each one of the contracted
basis set families by means of a small subset of the
GW100 dataset \cite{vanSetten:2015:5665,GW100GitHub}.
The subset is made exclusively of molecules containing
B, C, N, O, and/or F (the ccX family has
only been generated for these elements). The result is a
test set with 10 vertical ionization potentials 
(N$_2$, F$_2$, C$_4$, CF$_4$, BF, BN, 
CO$_2$, CO, O$_3$, C$_6$F$_6$)
and 6 vertical electron affinities corresponding to the
experimentally-bound anions
(F$_2$, C$_4$, BN, CO, O$_3$, C$_6$F$_6$). 
In order to directly compare
to existing benchmark data \cite{GW100GitHub,Govoni:2018:1895}, 
the results presented in this section used the ``pure'' PBE 
functional at the $G_0W_0$ level without basis set extrapolation.

Table \ref{tab:VIP} shows the vertical ionization potentials (VIP),
in eV, obtained with the different basis sets tested. It is
immediately apparent that the core-rich basis sets converge as
fast as the traditional cc-pV$n$Z and def2-$n$ZVP families. In contrast with the CLBEs case, the VIPs computed with all basis set families tend
to the same value.

\nocite{Trickl:1989:6006}
\nocite{VANLONKHUYZEN:1984:313}
\nocite{Ramanathan:1993:7838}
\nocite{BIERI:1981:281}
\nocite{Farber:1984:241}
\nocite{Eland:1977:5034}
\nocite{POTTS:1974:3}
\nocite{Katsumata:1984:1784}

\begin{table}[ht]
    \centering
    \caption{Vertical ionization potentials (in eV) obtained at the $G_0W_0$@PBE level of theory. Experimental results were taken from References \citenum{Trickl:1989:6006}--\!\citenum{Katsumata:1984:1784}.}
    \label{tab:VIP}
    \begin{tabular}{l c c c c c c c c c c}
        & N$_2$ & F$_2$ & C$_4$ & CF$_4$ & BF & BN & CO$_2$ & CO & O$_3$ & C$_6$F$_6$  \\\hline
        Exptl.    & 15.58 & 15.70 & 12.54 & 16.20 & 11.00 &       & 13.77 & 14.01 & 12.73 & 10.20 \\[2ex]
        def2-svp  & 14.48 & 14.39 & 10.40 & 14.67 & 10.19 & 10.67 & 12.61 & 13.15 & 11.50 & 8.77 \\
        def2-tzvp & 14.72 & 14.80 & 10.66 & 15.13 & 10.40 & 10.91 & 13.04 & 13.43 & 11.84 & 9.25 \\
        def2-qzvp & 14.89 & 14.96 & 10.78 & 15.37 & 10.56 & 11.01 & 13.25 & 13.57 & 11.96 & 9.49 \\[2ex]
        
        cc-pVDZ & 14.32 & 14.30 & 10.31 & 14.45 & 10.41 & 10.71 & 12.46 & 13.18 & 11.32 & 8.57 \\
        cc-pVTZ & 14.81 & 14.84 & 10.70 & 15.15 & 10.58 & 10.94 & 13.06 & 13.53 & 11.75 & 9.19 \\
        cc-pVQZ & 14.94 & 15.05 & 10.81 & 15.40 & 10.60 & 11.02 & 13.28 & 13.63 & 11.94 & 9.47 \\
        cc-pV5Z & 14.98 & 15.10 & 10.86 & 15.50 & 10.61 & 11.06 & 13.37 & 13.65 & 12.03 & 9.59 \\[2ex]
        
        pcSseg-1 & 14.56 & 14.58 & 10.46 & 14.89 & 10.45 & 10.82 & 12.80 & 13.37 & 10.90 & 9.04 \\
        pcSseg-2 & 14.73 & 14.79 & 10.67 & 15.18 & 10.42 & 10.92 & 13.06 & 13.45 & 11.90 & 9.36 \\
        pcSseg-3 & 14.92 & 14.94 & 10.81 & 15.38 & 10.55 & 11.02 & 13.27 & 13.59 & 11.98 & 9.54 \\
        pcSseg-4 & 14.96 & 15.03 & 10.86 & 15.48 & 10.56 & 11.05 & 13.36 & 13.64 & 12.03 & 9.61 \\[2ex]
        
        pcJ-1 & 14.56 & 14.65 & 10.49 & 14.97 & 10.51 & 10.85 & 12.86 & 13.42 & 11.69 & 9.09 \\
        pcJ-2 & 14.84 & 14.93 & 10.78 & 15.32 & 10.56 & 10.97 & 13.17 & 13.55 & 11.93 & 9.44 \\
        pcJ-3 & 14.93 & 14.99 & 10.82 & 15.42 & 10.53 & 11.03 & 13.30 & 13.60 & 12.00 & 9.56 \\
        pcJ-4 & 14.95 & 15.06 & 10.87 & 15.51 & 10.55 & 11.06 & 13.38 & 13.64 & 12.04 & 9.63 \\[2ex]
        
        ccX-DZ & 14.52 & 14.63 & 10.46 & 14.97 & 10.39 & 10.76 & 12.80 & 13.27 & 11.11 & 9.18 \\
        ccX-TZ & 14.78 & 14.84 & 10.71 & 15.25 & 10.48 & 10.94 & 13.13 & 13.49 & 11.92 & 9.42 \\
        ccX-QZ & 14.91 & 14.99 & 10.82 & 15.43 & 10.52 & 11.03 & 13.30 & 13.60 & 12.00 & 9.56 \\
        ccX-5Z & 14.96 & 15.07 & 10.87 & 15.52 & 10.54 & 11.06 & 13.38 & 13.64 & 12.04 & 9.63 \\\hline 
    \end{tabular}
\end{table}

Slightly larger variations in the rate of convergence for each 
basis set family can be seen in the vertical electron affinities (VEA)
series shown in Table \ref{tab:VEA}. Here, it is evident that the
core-rich basis sets offer better VEAs at the double- and triple-$\zeta$
level than the corresponding cc-pV$n$Z and def2-$n$zvp sets, especially
for molecules containing fluorine atoms. The good performance of the
core-rich basis sets for VEAs is partly explained by the presence
of some more diffuse basis functions. Table \ref{tab:diffuse} shows
the smallest exponent of each angular momentum for all double-$\zeta$
basis sets. To facilitate, all exponents are given as the fraction
of the corresponding def2-SVP ones. It is evident that the ccX-$n$Z
family is consistently more diffuse than their counterparts, hence
it good performance for $GW$ VEAs calculations, but the story is 
not as evident for the other core-rich families. The remaining 
difference is explained by the additional variation flexibility
core-rich basis set have in comparison to standard ones.
For example, the uncontracted def2-SVP VEAs for F$_2$ and 
C$_6$F$_6$ (-1.40 eV and -1.45 eV, respectively) show 
a slight improvement toward the basis set limit with respect
to the standard def2-SVP.

It is reassuring to see that the core-rich basis sets are
very well balanced to describe both core- and valence-excitations
at the $GW$ level. 

\nocite{Ayala:1981:768}
\nocite{Arnold:1991:8753}
\nocite{Arnold:1994:912}
\nocite{Eustis:2007:114312}

\begin{table}[ht]
    \centering
    \caption{Vertical electron affinities (in eV) obtained at the $G_0W_0$@PBE level of theory. Experimental values taken from References \citenum{Ayala:1981:768}-\citenum{Eustis:2007:114312}.}
    \label{tab:VEA}
    \begin{tabular}{l c c c c c c}
                  & F$_2$ & C$_4$ &  BN   &  CO   & O$_3$ & C$_6$F$_6$  \\\hline
        Exptl.    &  1.24 & 3.88  &  3.16 &  1.33 & 2.10 & 0.70  \\[2ex]
        def2-svp  & -1.61 & 2.14  &  3.49 & -2.00 & 0.74 & -1.54 \\
        def2-tzvp &  0.17 & 2.72  &  3.81 & -0.97 & 1.89 & -0.96 \\
        def2-qzvp &  0.70 & 2.94  &  3.95 & -0.67 & 2.30 & -0.66 \\[2ex]
        
        cc-pVDZ   & -1.63 & 1.95  &  3.27 & -2.23 & 0.51 & -1.73 \\
        cc-pVTZ   & -0.23 & 2.63  &  3.72 & -1.19 & 1.62 & -1.32 \\
        cc-pVQZ   &  0.43 & 2.90  &  3.92 & -0.78 & 2.13 & -0.78 \\
        cc-pV5Z   &  0.87 & 3.03  &  4.01 & -0.55 & 2.45 & -0.48 \\[2ex]
        
        pcSseg-1  & -0.88 & 2.23  &  3.49 & -1.74 & 1.13 & -1.19 \\
        pcSseg-2  &  0.38 & 2.77  &  3.84 & -0.88 & 2.07 & -0.84 \\
        pcSseg-3  &  0.85 & 2.98  &  3.99 & -0.59 & 2.42 & -0.55 \\
        pcSseg-4  &  1.00 & 3.06  &  4.04 & -0.49 & 2.56 & -0.26 \\[2ex]
        
        pcJ-1     & -0.80 & 2.24  &  3.51 & -1.72 & 1.18 & -1.13 \\
        pcJ-2     &  0.53 & 2.81  &  3.88 & -0.83 & 2.17 & -0.91 \\
        pcJ-3     &  0.93 & 3.00  &  4.01 & -0.55 & 2.48 & -0.43 \\
        pcJ-4     &  1.04 & 3.07  &  4.04 & -0.47 & 2.59 & -0.21 \\[2ex]
        
        ccX-DZ    &  0.36 & 2.52  &  3.76 & -1.15 & 1.90 & -1.08 \\
        ccX-TZ    &  0.73 & 2.86  &  3.92 & -0.72 & 2.28 & -0.62 \\
        ccX-QZ    &  0.95 & 3.01  &  4.01 & -0.55 & 2.49 & -0.37 \\
        ccX-5Z    &  1.05 & 3.07  &  4.04 & -0.47 & 2.59 & -0.22 \\\hline 
    \end{tabular}
\end{table}

\begin{table}[]
    \centering
    \caption{Diffuse exponents present in each double-$\zeta$ basis set tested as a fraction of the corresponding def2-svp exponent.}
    \label{tab:diffuse}
    \begin{tabular}{l c c c c c c c}
              & \multicolumn{3}{c}{Boron} & & \multicolumn{3}{c}{Carbon}\\\cline{2-4}\cline{6-8}
              & S     & P     & D         & & S    & P     & D   \\\hline
    def2-svp  & 1.00  & 1.00  & 1.00      & & 1.00 & 1.00 & 1.00 \\ 
    cc-pVDZ   & 1.25  & 0.99  & 0.69      & & 1.22 & 0.99 & 0.69 \\ 
    pcSseg-1  & 1.19  & 0.90  & 1.40      & & 1.14 & 0.89 & 1.00 \\ 
    pcJ-1     & 1.27  & 0.90  & 1.40      & & 1.23 & 0.89 & 1.00 \\ 
    ccX-DZ    & 0.89  & 0.52  & 0.69      & & 0.85 & 0.50 & 0.55 \\[2ex]

              & \multicolumn{3}{c}{Nitrogen} & & \multicolumn{3}{c}{Oxygen}\\\cline{2-4}\cline{6-8}
              & S     & P     & D         & & S    & P     & D   \\\hline
    def2-svp  & 1.00  & 1.00  & 1.00      & & 1.00 & 1.00 & 1.00 \\ 
    cc-pVDZ   & 1.20  & 1.00  & 0.82      & & 1.18 & 1.00 & 0.99 \\ 
    pcSseg-1  & 1.12  & 0.89  & 0.90      & & 1.12 & 0.89 & 0.83 \\ 
    pcJ-1     & 1.22  & 0.89  & 0.90      & & 1.20 & 0.89 & 0.83 \\ 
    ccX-DZ    & 0.83  & 0.49  & 0.82      & & 0.81 & 0.44 & 0.99 \\[2ex]
    
              & \multicolumn{3}{c}{Fluorine} & & \multicolumn{3}{c}{}\\\cline{2-4}
              & S     & P     & D         \\\cline{1-4}
    def2-svp  & 1.00  & 1.00  & 1.00      \\ 
    cc-pVDZ   & 1.17  & 1.00  & 1.17      \\ 
    pcSseg-1  & 1.12  & 0.89  & 0.79      \\ 
    pcJ-1     & 1.20  & 0.89  & 0.79      \\ 
    ccX-DZ    & 0.80  & 0.45  & 1.17      \\\cline{1-4}
    \end{tabular}
\end{table}

\section{Summary}
\label{summ}
Convergence of the core-level binding energies obtained
in $GW$ calculations have often been deemed hard to achieve  
with respect to the basis set quality. Here, we have shown that
the def2-$n$zvp and cc-pV$n$Z basis sets families are shown to have 
substantial contraction errors when used to compute $GW$ core-level
binding energies of molecules containing first row atoms. Furthermore, we have shown that basis set extrapolations 
using such bases might be unreliable, as in the case of the def2-$n$zvp
family, or might yield underestimates due to the very slow convergence
seen with the cc-pV$n$Z family.

Exposing all the core basis functions by uncontracting either basis
set adds enough flexibility to the calculation, shifting the binding energies
to higher values and showing much faster convergence behavior. The same
behavior can be observed with partially uncontracted basis sets,
like pcJ-$n$, pcSseg-$n$, and ccX-$n$ families. In particular, 
both pcJ-$n$ and ccX-$n$ seem very well suited to study $GW$
core-level binding energies.

The quality of $GW$ valence charged excitations does not degrade
by using one of the core-rich basis sets. In fact, vertical 
electron affinities achieve faster convergence rates. 

We therefore recommend the use of either pcJ-$n$ or ccX-$n$Z families
for core-level and valence $GW$ calculations. 
However, it is important to note that both
basis sets do impact the speed of the calculation. 
A more cost-effective choice might be the use of an
uncontracted standard basis set,
like cc-pV$n$Z and def2-$n$zvp. Note that standard J, JK, an RI 
fitting bases are not recommended when using uncontracted basis sets.\\
Another fact worth noting is that the current study focused only on the
elements C to F (present in the CORE65 database); therefore the conclusions
of this paper might not necessarily hold when applying GWA to molecules containing heavier elements.

\begin{acknowledgement}

The authors acknowledge funding from the Center for Scalable and
Predictive methods for Excitation and Correlated phenomena (SPEC),
which is funded by the U.S. Department of Energy (DOE), Office of
Science, Office of Basic Energy Sciences, the Division of Chemical
Sciences, Geosciences, and Biosciences. This research also benefited
from computational resources provided by EMSL, a DOE Office of Science
User Facility sponsored by the Office of Biological and Environmental
Research and located at the Pacific Northwest National Laboratory
(PNNL). PNNL is operated by Battelle Memorial Institute for the United
States Department of Energy under DOE contract number
DE-AC05-76RL1830. This research used resources of the National Energy
Research Scientific Computing Center (NERSC), a U.S. Department of
Energy Office of Science User Facility located at Lawrence Berkeley
National Laboratory, operated under Contract No. DE-AC02-05CH11231.

\end{acknowledgement}

\begin{suppinfo}
Tables with all individual core-level binding energies (in eV) used to make the plots shown in Figures 1--5.
Comparison of mean errors and mean absolute errors with respect to experiment.
\end{suppinfo}

\bibliography{gw.bib}

\end{document}


\renewcommand*{\arraystretch}{0.9}

\begin{center}
\rowcolors{2}{gray!25}{white}
\begin{table}
\captionsetup{font=scriptsize}
\centering
\caption{Core-level binding energies [eV] for the contracted def2-$n$zvp basis set family. The headers DZ, TZ, and QZ correspond to def2-svp, def2-tzvp, and def2-qzvp, respectively.}
{\scriptsize
\begin{tabularx}{\textwidth}{l l X X X X X}
\rowcolor{white} name & core level & & Exptl. & DZ & TZ & QZ  \\\hline
methane              & C          & & 290.84 & 291.50 & 290.43 & 290.28 \\
ethane               & C          & & 290.71 & 291.42 & 290.38 & 290.26 \\
ethene               & C          & & 290.82 & 291.65 & 290.53 & 290.48 \\
ethyne               & C          & & 291.25 & 292.15 & 290.93 & 290.74 \\
carbon monoxide      & O          & & 542.10 & 542.87 & 541.36 & 541.04 \\
carbon monoxide      & C          & & 296.23 & 296.98 & 295.61 & 295.36 \\
carbon dioxide       & O          & & 541.32 & 541.83 & 540.48 & 540.16 \\
\hline
\end{tabularx}
}
\end{table}

\rowcolors{2}{gray!25}{white}
\begin{table}
\captionsetup{font=scriptsize}
\centering
\caption{\footnotesize Core-level binding energies [eV] for the uncontracted un-def2-$n$zvp basis set family. The headers DZ, TZ, and QZ correspond to un-def2-svp, un-def2-tzvp, and un-def2-qzvp, respectively.}
{\scriptsize
\begin{tabularx}{\textwidth}{l l X X X X X}
\rowcolor{white} name & core level & & Exptl. & DZ & TZ & QZ  \\\hline
methane              & C          &  & 290.84 & 290.89 & 290.86 & 290.93 \\
ethane               & C          &  & 290.71 & 290.87 & 290.84 & 290.92 \\
ethene               & C          &  & 290.82 & 291.05 & 290.97 & 291.05 \\
ethyne               & C          &  & 291.25 & 291.49 & 291.31 & 291.36 \\
carbon monoxide      & O          &  & 542.10 & 541.96 & 541.64 & 541.66 \\
carbon monoxide      & C          &  & 296.23 & 296.40 & 296.07 & 296.09 \\
carbon dioxide       & O          &  & 541.32 & 541.08 & 540.74 & 540.78 \\
\hline
\end{tabularx}
}
\end{table}

\rowcolors{2}{gray!25}{white}
\begin{table}
\captionsetup{font=scriptsize}
\centering
\caption{\footnotesize Core-level binding energies [eV] for the contracted cc-pV$n$Z basis set family. The headers DZ, TZ, QZ, and 5Z correspond to cc-pVDZ, cc-pVTZ, cc-pVQZ, and cc-pV5Z, respectively.}
{\scriptsize
\begin{tabularx}{\textwidth}{l l X X X X X X}
\rowcolor{white} name & core level & & Exptl. & DZ & TZ & QZ & 5Z  \\\hline
methane              & C          &  & 290.84 & 291.93 & 290.20 & 290.42 & 290.52 \\
ethane               & C          &  & 290.71 & 291.84 & 290.17 & 290.39 & 290.46 \\
ethene               & C          &  & 290.82 & 292.01 & 290.34 & 290.54 & 290.64 \\
ethyne               & C          &  & 291.25 & 292.50 & 290.83 & 290.89 & 290.92 \\
carbon monoxide      & O          &  & 542.10 & 543.19 & 541.19 & 541.17 & 541.23 \\
carbon monoxide      & C          &  & 296.23 & 297.58 & 295.51 & 295.58 & 295.57 \\
carbon dioxide       & O          &  & 541.32 & 542.14 & 540.24 & 540.30 & 540.32 \\
\hline
\end{tabularx}
}
\end{table}

\rowcolors{2}{gray!25}{white}
\begin{table}
\captionsetup{font=scriptsize}
\centering
\caption{\footnotesize Core-level binding energies [eV] for the uncontracted un-cc-pV$n$Z basis set family. The headers DZ, TZ, QZ, and 5Z correspond to un-cc-pVDZ, un-cc-pVTZ, un-cc-pVQZ, and un-cc-pV5Z, respectively.}
{\scriptsize
\begin{tabularx}{\textwidth}{l l X X X X X X}
\rowcolor{white} name & core level & & Exptl. & DZ & TZ & QZ & 5Z  \\\hline
methane              & C          &  & 290.84 & 290.95 & 290.91 & 290.94 & 290.98 \\
ethane               & C          &  & 290.71 & 290.92 & 290.90 & 290.91 & 290.96 \\
ethene               & C          &  & 290.82 & 291.08 & 290.99 & 291.04 & 291.09 \\
ethyne               & C          &  & 291.25 & 291.50 & 291.40 & 291.42 & 291.39 \\
carbon monoxide      & O          &  & 542.10 & 541.93 & 541.74 & 541.70 & 541.65 \\
carbon monoxide      & C          &  & 296.23 & 296.37 & 296.27 & 296.15 & 296.12 \\
carbon dioxide       & O          &  & 541.32 & 540.88 & 540.79 & 540.82 & 540.76 \\
\hline
\end{tabularx}
}
\end{table}

\rowcolors{2}{gray!25}{white}
\begin{table}
\captionsetup{font=scriptsize}
\centering
\caption{\footnotesize Core-level binding energies [eV] for the contracted pcSseg-$n$ basis set family. The headers DZ, TZ, and QZ correspond to pcSseg-1, pcSseg-2, and pcSseg-3, respectively.}
{\scriptsize
\begin{tabularx}{\textwidth}{l l X X X X X}
\rowcolor{white} name & core level & & Exptl. & DZ & TZ & QZ  \\\hline
methane              & C          &  & 290.84 & 291.12 & 291.39 & 291.12 \\
ethane               & C          &  & 290.71 & 291.11 & 291.31 & 291.03 \\
ethene               & C          &  & 290.82 & 291.29 & 291.42 & 291.12 \\
ethyne               & C          &  & 291.25 & 291.69 & 291.82 & 291.46 \\
carbon monoxide      & O          &  & 542.10 & 542.12 & 542.29 & 541.77 \\
carbon monoxide      & C          &  & 296.23 & 296.75 & 296.85 & 296.19 \\
carbon dioxide       & O          &  & 541.32 & 541.21 & 541.38 & 540.87 \\
\hline
\end{tabularx}
}
\end{table}

\rowcolors{2}{gray!25}{white}
\begin{table}
\captionsetup{font=scriptsize}
\centering
\caption{\footnotesize Core-level binding energies [eV] for the uncontracted un-pcSseg-$n$ basis set family. The headers DZ, TZ, and QZ correspond to un-pcSseg-1, un-pcSseg-2, and un-pcSseg-3, respectively.}
{\scriptsize
\begin{tabularx}{\textwidth}{l l X X X X X}
\rowcolor{white} name & core level & & Exptl. & DZ & TZ & QZ  \\\hline
methane              & C          &  & 290.84 & 290.89 & 290.89 & 290.97 \\
ethane               & C          &  & 290.71 & 290.88 & 290.81 & 290.96 \\
ethene               & C          &  & 290.82 & 291.07 & 291.01 & 291.04 \\
ethyne               & C          &  & 291.25 & 291.54 & 291.39 & 291.39 \\
carbon monoxide      & O          &  & 542.10 & 541.78 & 541.64 & 541.64 \\
carbon monoxide      & C          &  & 296.23 & 296.40 & 296.11 & 296.11 \\
carbon dioxide       & O          &  & 541.32 & 540.88 & 540.75 & 540.75 \\
\hline
\end{tabularx}
}
\end{table}

\rowcolors{2}{gray!25}{white}
\begin{table}
\captionsetup{font=scriptsize}
\centering
\caption{\footnotesize Core-level binding energies [eV] for the contracted pcJ-$n$ basis set family.  The headers DZ, TZ, and QZ correspond to pcJ-1, pcJ-2, and pcJ-3, respectively.}
{\scriptsize
\begin{tabularx}{\textwidth}{l l X X X X X}
\rowcolor{white} name & core level & & Exptl. & DZ & TZ & QZ  \\\hline
methane              & C          &  & 290.84 & 290.93 & 290.82 & 291.00 \\
ethane               & C          &  & 290.71 & 290.91 & 290.80 & 290.98 \\
ethene               & C          &  & 290.82 & 291.11 & 290.92 & 291.07 \\
ethyne               & C          &  & 291.25 & 291.56 & 291.27 & 291.38 \\
carbon monoxide      & O          &  & 542.10 & 541.90 & 541.53 & 541.64 \\
carbon monoxide      & C          &  & 296.23 & 296.44 & 296.07 & 296.11 \\
carbon dioxide       & O          &  & 541.32 & 541.00 & 540.64 & 540.75 \\
\hline
\end{tabularx}
}
\end{table}

\rowcolors{2}{gray!25}{white}
\begin{table}
\captionsetup{font=scriptsize}
\centering
\caption{\footnotesize Core-level binding energies [eV] for the uncontracted un-pcJ-$n$ basis set family.  The headers DZ, TZ, and QZ correspond to un-pcJ-1, un-pcJ-2, and un-pcJ-3, respectively.}
{\scriptsize
\begin{tabularx}{\textwidth}{l l X X X X X}
\rowcolor{white} name & core level & & Exptl. & DZ & TZ & QZ  \\\hline
methane              & C          &  & 290.84 & 291.08 & 291.02 & 291.07 \\
ethane               & C          &  & 290.71 & 291.06 & 291.00 & 291.05 \\
ethene               & C          &  & 290.82 & 291.25 & 291.11 & 291.12 \\
ethyne               & C          &  & 291.25 & 291.70 & 291.47 & 291.44 \\
carbon monoxide      & O          &  & 542.10 & 541.94 & 541.73 & 541.73 \\
carbon monoxide      & C          &  & 296.23 & 296.56 & 296.25 & 296.19 \\
carbon dioxide       & O          &  & 541.32 & 541.05 & 540.83 & 540.83 \\
\hline
\end{tabularx}
}
\end{table}

\rowcolors{2}{gray!25}{white}
\begin{table}
\captionsetup{font=scriptsize}
\centering
\caption{\footnotesize Core-level binding energies [eV] for the contracted ccX-$n$Z basis set family.  The headers DZ, TZ, and QZ correspond to ccX-DZ, ccX-TZ, and ccX-QZ, respectively.}
{\scriptsize
\begin{tabularx}{\textwidth}{l l X X X X X}
\rowcolor{white} name & core level & & Exptl. & DZ & TZ & QZ  \\\hline
methane              & C          &  & 290.84 & 291.08 & 291.00 & 291.06 \\
ethane               & C          &  & 290.71 & 291.06 & 290.99 & 291.05 \\
ethene               & C          &  & 290.82 & 291.17 & 291.07 & 291.12 \\
ethyne               & C          &  & 291.25 & 291.45 & 291.41 & 291.45 \\
carbon monoxide      & O          &  & 542.10 & 541.78 & 541.92 & 541.76 \\
carbon monoxide      & C          &  & 296.23 & 296.35 & 296.14 & 296.17 \\
carbon dioxide       & O          &  & 541.32 & 540.85 & 540.90 & 540.97 \\
\hline
\end{tabularx}
}
\end{table}

\rowcolors{2}{gray!25}{white}
\begin{table}
\captionsetup{font=scriptsize}
\centering
\caption{\footnotesize Core-level binding energies [eV] for the uncontracted un-ccX-$n$Z basis set family. The headers DZ, TZ, and QZ correspond to un-ccX-DZ, un-ccX-TZ, and un-ccX-QZ, respectively.}
{\scriptsize
\begin{tabularx}{\textwidth}{l l X X X X X}
\rowcolor{white} name & core level & & Exptl. & DZ & TZ & QZ  \\\hline
methane              & C          &  & 290.84 & 291.05 & 291.08 & 291.14 \\
ethane               & C          &  & 290.71 & 291.04 & 291.06 & 291.13 \\
ethene               & C          &  & 290.82 & 291.16 & 291.14 & 291.19 \\
ethyne               & C          &  & 291.25 & 291.45 & 291.49 & 291.55 \\
carbon monoxide      & O          &  & 542.10 & 541.81 & 541.91 & 541.86 \\
carbon monoxide      & C          &  & 296.23 & 296.36 & 296.21 & 296.27 \\
carbon dioxide       & O          &  & 541.32 & 540.87 & 540.97 & 540.97 \\
\hline
\end{tabularx}
}
\end{table}
\end{center}

\begin{center}
\renewcommand*{\arraystretch}{1.1}
\begin{table}
\caption{Mean errors (ME) and mean absolute errors (MAE), with respect to experiment, for the core-level binding energies obtained with contracted basis sets and with relativistic corrections. All values in eV.}
\begin{tabular}{l S[table-format=3.2] S[table-format=3.2] S[table-format=3.2] S[table-format=3.2] c S[table-format=3.2] S[table-format=3.2] S[table-format=3.2] S[table-format=3.2]}
          & \multicolumn{4}{c}{ME} & & \multicolumn{4}{c}{MAE} \\\cline{2-5}\cline{7-10}
          & C & N & O & F & \phantom{A} & C & N & O & F\\\hline
def2-svp  & 0.97 & 1.01  & 0.94 & 0.88 & & 0.97 & 1.01 & 0.94 & 0.88 \\
def2-tzvp & -0.11 & -0.16 & -0.29 & -0.53 & & 0.14 & 0.25 & 0.38 & 0.53 \\
def2-qzvp & -0.21 & -0.41 & -0.62 & -1.0 & & 0.22 & 0.49 & 0.68 & 1.00 \\[2ex]

cc-pVDZ & 1.31 & 1.37 & 1.31 & 1.32 & & 1.31 & 1.37 & 1.31 & 1.32 \\
cc-pVTZ & -0.30 & -0.42 & -0.55 & -0.81 & & 0.30 & 0.48 & 0.62 & 0.81 \\
cc-pVQZ & -0.08 & -0.25 & -0.49 & -0.85 & & 0.14 & 0.35 & 0.57 & 0.85 \\
cc-pV5Z & -0.01 & -0.15 & -0.46 & -0.90 & & 0.14 & 0.25 & 0.54 & 0.90 \\[2ex]

pcSseg-1 & 0.74 & 0.58 & 0.37 & 0.19 & & 0.74 & 0.58 & 0.37 & 0.19 \\
pcSseg-2 & 0.84 & 0.75 & 0.59 & 0.37 & & 0.84 & 0.75 & 0.59 & 0.37 \\
pcSseg-3 & 0.55 & 0.34 & 0.06 & -0.35 & & 0.55 & 0.34 & 0.19 & 0.35 \\[2ex]

pcJ-1 & 0.54 & 0.37 & 0.16 & -0.05 & & 0.54 & 0.37 & 0.20 & 0.11 \\
pcJ-2 & 0.35 & 0.13 & -0.15 & -0.52 & & 0.35 & 0.19 & 0.28 & 0.52 \\
pcJ-3 & 0.50 & 0.24 & -0.06 & -0.48 & & 0.50 & 0.25 & 0.22 & 0.48 \\[2ex]

ccX-DZ & 0.65 & 0.35 & 0.06 & -0.31 & & 0.65 & 0.35 & 0.18 & 0.31 \\
ccX-TZ & 0.51 & 0.24 & 0.08 & -0.29 & & 0.51 & 0.25 & 0.20 & 0.29 \\
ccX-QZ & 0.56 & 0.32 & 0.06 & -0.30 & & 0.56 & 0.32 & 0.19 & 0.30 \\[2ex]

Golze et al.\cite{Golze:2020:1840} & 0.19 & -0.01 & -0.40 & -0.83 & & 0.24 & 0.16 & 0.48 & 0.83 \\\hline
\hline
\end{tabular}
\end{table}

\begin{table}
\caption{Mean errors (ME) and mean absolute errors (MAE), with respect to experiment, for the core-level binding energies obtained with uncontracted basis sets and with relativistic corrections. All values in eV.}
\begin{tabular}{l S[table-format=3.2] S[table-format=3.2] S[table-format=3.2] S[table-format=3.2] c S[table-format=3.2] S[table-format=3.2] S[table-format=3.2] S[table-format=3.2]}
          & \multicolumn{4}{c}{ME} & & \multicolumn{4}{c}{MAE} \\\cline{2-5}\cline{7-10}
          & C & N & O & F & \phantom{A} & C & N & O & F\\\hline
un-def2-svp  & 0.46 & 0.39 & 0.18 & -0.18 & & 0.46 & 0.39 & 0.26 & 0.18 \\
un-def2-tzvp & 0.40 & 0.18 & -0.05 & -0.44 & & 0.41 & 0.20 & 0.22 & 0.44 \\
un-def2-qzvp & 0.44 & 0.22 & 0.03 & -0.39 & & 0.44 & 0.24 & 0.24 & 0.39 \\[2ex]

un-cc-pVDZ & 0.43 & 0.28 & 0.07 & -0.10 & & 0.43 & 0.28 & 0.23 & 0.11 \\
un-cc-pVTZ & 0.40 & 0.24 & -0.03 & -0.36 & & 0.40 & 0.24 & 0.22 & 0.36 \\
un-cc-pVQZ & 0.46 & 0.30 & 0.01 & -0.34 & & 0.46 & 0.31 & 0.19 & 0.34 \\
un-cc-pV5Z & 0.47 & 0.24 & -0.04 & -0.41 & & 0.47 & 0.25 & 0.21 & 0.41 \\[2ex]

un-pcSseg-1 & 0.52 & 0.28 & 0.04 & -0.21 & & 0.52 & 0.29 & 0.19 & 0.21 \\
un-pcSseg-2 & 0.44 & 0.23 & -0.04 & -0.40 & & 0.44 & 0.24 & 0.21 & 0.40 \\
un-pcSseg-3 & 0.48 & 0.22 & -0.06 & -0.50 & & 0.48 & 0.23 & 0.22 & 0.50 \\[2ex]

un-pcJ-1 & 0.68 & 0.45 & 0.21 & -0.06 & & 0.68 & 0.45 & 0.23 & 0.11 \\
un-pcJ-2 & 0.55 & 0.31 & 0.04 & -0.30 & & 0.55 & 0.31 & 0.19 & 0.30 \\
un-pcJ-3 & 0.56 & 0.32 & 0.03 & -0.40 & & 0.56 & 0.32 & 0.20 & 0.40 \\[2ex]

un-ccX-DZ & 0.64 & 0.35 & 0.07 & -0.30 & & 0.64 & 0.35 & 0.18 & 0.30 \\
un-ccX-TZ & 0.59 & 0.32 & 0.15 & -0.22 & & 0.59 & 0.32 & 0.22 & 0.22 \\
un-ccX-QZ & 0.65 & 0.41 & 0.15 & -0.19 & & 0.65 & 0.41 & 0.21 & 0.19 \\[2ex]

Golze et al.\cite{Golze:2020:1840} & 0.19 & -0.01 & -0.40 & -0.83 & & 0.24 & 0.16 & 0.48 & 0.83 \\\hline
\end{tabular}
\end{table}
\end{center}